%% file: 0_main.tex
\documentclass[conference]{IEEEtran}

\usepackage[pdftex]{graphicx}
\usepackage{amsmath}

\usepackage[caption=false,font=footnotesize]{subfig}

\usepackage{url}

\usepackage{todonotes}
\usepackage{gensymb}
\usepackage{booktabs}
\usepackage{multirow}
\usepackage{hyperref}
\usepackage{cleveref}

\newcommand\copyrighttext{%
  \footnotesize \textcopyright 2021 IEEE. Personal use of this material is permitted.
  Permission from IEEE must be obtained for all other uses, in any current or future
  media, including reprinting/republishing this material for advertising or promotional
  purposes, creating new collective works, for resale or redistribution to servers or
  lists, or reuse of any copyrighted component of this work in other works.
  DOI: \href{https://doi.org/10.1109/ICC42927.2021.9500958}{10.1109/ICC42927.2021.9500958}}
\newcommand\copyrightnotice{%
\begin{tikzpicture}[remember picture,overlay]
\node[anchor=south,yshift=10pt] at (current page.south) {\fbox{\parbox{\dimexpr\textwidth-\fboxsep-\fboxrule\relax}{\copyrighttext}}};
\end{tikzpicture}%
}

\usepackage{glossaries} 
\setacronymstyle{long-short}

\include{5_acronyms}

\begin{document}

\title{Keys from the Sky: A First Exploration of Physical-Layer Security Using Satellite Links}

\author{
\IEEEauthorblockN{Pascal Zimmer\textsuperscript{2}, Roland Weinreich\textsuperscript{1},  Christian T. Zenger\textsuperscript{1,3}, Aydin Sezgin\textsuperscript{1}, Christof Paar\textsuperscript{1,2}}\\
\IEEEauthorblockA{\textsuperscript{1}Ruhr University Bochum, Horst Görtz Institute for IT-Security, Bochum, Germany}
\IEEEauthorblockA{\textsuperscript{2}Max Planck Institute for Security and Privacy, Bochum, Germany}
\IEEEauthorblockA{\textsuperscript{3}PHYSEC GmbH, Bochum, Germany}
\IEEEauthorblockA{E-Mail: \{roland.weinreich,christian.zenger,aydin.sezgin\}@rub.de, \{pascal.zimmer,christof.paar\}@csp.mpg.de}
}

\maketitle

\copyrightnotice

\begin{abstract}
In this paper, we investigate \gls{PLS} methods for proximity-based group-key establishment and proof of location. Fields of application include secure car-to-car communication, privacy-preserving and secure distance evidence for healthcare or location-based feature activation. Existing technologies do not solve the problem satisfactorily, due to communication restrictions, e.g., \gls{UWB} based time of flight measurements, or trusted hardware, e.g., using \gls{GNSS} positioning data. 

We introduce \gls{PLS} as a solution candidate. It is information theoretically secure, which also means post-quantum resistant, and has the potential to run on resource constrained devices with low latency. Furthermore, we use wireless channel properties of satellite-to-Earth links, demonstrate the first feasibility study using off-the-shelf hardware testbeds and present first evaluation results and future directions for research.
\end{abstract}

\IEEEpeerreviewmaketitle

 
\input{1_intro}

\input{2_background}

\input{3_results}

\input{4_discussion_future_work}


\input{6_bibliography}

\end{document}

%% file: 5_acronyms.tex
\newacronym{GNSS}{GNSS}{global navigation satellite system}%
\newacronym{UWB}{UWB}{ultra-wide band}%
\newacronym{IPP}{IPP}{ionospheric pierce points}
\newacronym{IoT}{IoT}{Internet of Things}
\newacronym{LBAC}{LBAC}{location-based access-control systems}
\newacronym{LBS}{LBS}{location-based services}
\newacronym{MEO}{MEO}{Medium Earth Orbit}
\newacronym{PLS}{PLS}{physical-layer security}%

%% file: 1_intro.tex
\section{Introduction}

\noindent Mutual cryptographic secrets are a central requirement for virtually any security solution, ranging from the \gls{IoT} to classical internet applications. Traditional public-key based solutions for key establishment have the drawback of a complex public-key infrastructure and high computational complexity. Authenticated key establishment based on context, location or proximity is an attractive alternative in many application scenarios. This holds in particular in situations where devices have never interacted with each other before but are placed in proximity. We note that this is often given in IoT and other emerging applications.

This includes safety-sensitive systems, such as V2X infrastructures or autonomous driving, where manipulations of the communication can have catastrophic consequences. Other application require privacy-preserving and anonymous communication, such as \gls{LBS} that offer services based on the user's location. Moreover, they can also be used as \gls{LBAC} for location-based feature activation or limiting communication to larger areas, such as cities, towns, and factories.

Unfortunately, today's approaches are either limited to short distances, e.g.\ pairwise distance bounding with \gls{UWB} based time of flight measurements, or they force us to trust a third authority. Trusted authorities, such as authentication servers create security bottlenecks within protocols, due to possible compromise or unavailability.

The idea of \textit{keys from the sky} is motivated by Mathur et al.~\cite{mathur_proximate_2011}, who presented the initial \gls{PLS} concept of channel diversity based authentication and by Zenger et al.~\cite{zenger_authenticated_2016}, who presented a vicinity-based authenticated key establishment scheme. A scenario of long-range communication with low data rates and high packet loss has also been successfully used for key-derivation in LoRa(WAN)-based systems \cite{xu_lora-key_2019, ruotsalainen_experimental_2020}. The paper at hand investigates for the first time Mathur's approach using satellite-to-Earth radio links. It tries to answer the question whether proximity based group-key establishment schemes that intrinsically prove the relative location of group members are feasible. This will also answer whether correlated channel observations by two or more parties  enable them to prove their proximity to each other. 

\begin{figure}[t!]
	\centering
	\includegraphics[width=\columnwidth-0.5cm]{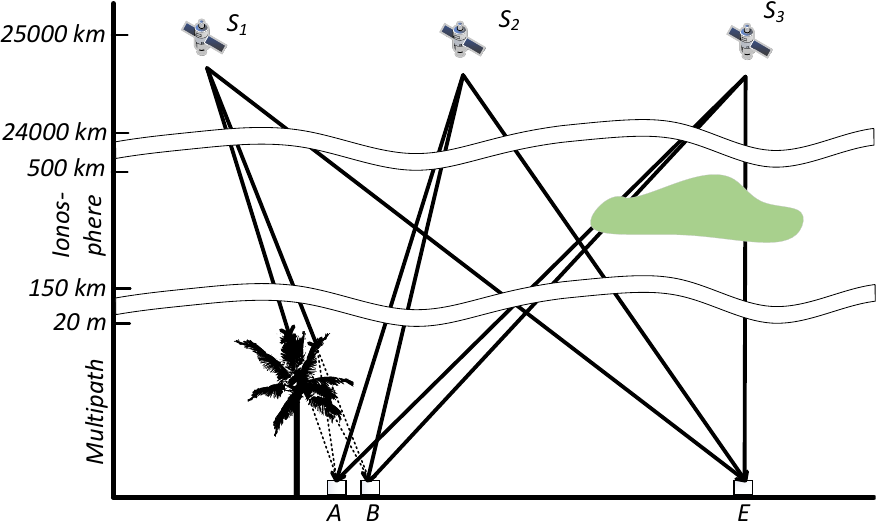}
	\caption{Our system model: utilizing spatial diversity of \gls{GNSS} satellite-to Earth channel for proximity-based key extraction. Not true to scale.}
	\label{fig:system_model_bidirectional}
\end{figure}

The communication scenario is shown in \Cref{fig:system_model_bidirectional}, highlighting a uni-directional setting in which a trustworthy \gls{GNSS} satellite is broadcasting signals that are picked up almost simultaneously by multiple receivers on Earth. This signal exhibits amplitude, phase and runtime changes due to clouds of ionized particles and multipath effects. Heavy multipath propagation is assumed to emerge, for example, if the satellite is close to the horizon and signals propagate through obstructions and other objects close to Earth's surface. 

Based on the physical property \emph{spatial diversity} of wireless channels, we create the hypothesis that two parties (Alice $A$, Bob $B$) in proximity to each other, probing the channel simultaneously, are able to obtain correlating channel observations from ambient signals of a satellite $S$. On the other hand, the attacker (Eve $E$) that is sufficiently far away, is not able to obtain correlating channel observations as the signals are experiencing different physical variations. Based on the channel observations, parties in proximity are able to extract a secret key. Pairwise or group-wise key establishment schemes known from the literature, e.g., \cite{mathur_proximate_2011, zenger_authenticated_2016} and \cite{truyen_thai_secret_2015, jiao_efficient_2020}, are candidates which can be build on top of our method.

The mutual awareness of the physical proximity is used for authentication. This also holds, if the legitimate parties in close physical proximity have never interacted with each other before or do not share any cryptographic security parameters. They are interested in exchanging confidential information, whereas an attacker outside a defined zone should not be able to obtain the key and correctly decrypt this communication. 

Due to the fact that the \gls{IoT} is growing at a rapid pace with a vast number of devices that are equipped with satellite receivers, our approach might have a significant impact for future decentralized trust establishment schemes. Furthermore, our real-world approach is strengthened by the fact that there are 4 independent \gls{GNSS} constellations that consist of a total of 128 satellites. Hence, our approach has the potential to create a secure multiparty basis of trust using 4 (possibly semi-trustworthy) instances. 

The main contributions of this paper are:
\begin{enumerate}
    \item We introduced the idea of a global multiparty-based and privacy-preserving proximity verification method.  
    \item We implement a proof-of-concept system based on low-cost off-the-shelf hardware and real-world requirements.
    \item We present the first empirical study employing \gls{PLS} on \gls{GNSS} satellite links: the general feasibility and first detailed evaluation results of our approach are shown, including an analysis and discussion of criteria for selecting useful satellite channel conditions.
\end{enumerate}

%% file: 2_background.tex
\section{Methodology and Fundamentals}

The paper at hand pursues a practice-oriented approach for designing a secure proximity-based key establishment protocol based on the idea of Mathur et al.~\cite{mathur_proximate_2011}. To make the approach real-world applicable, we consider three aspects more closely:
\begin{enumerate}
    \item Global availability.
    \item No trusted third party.
    \item Physical-layer properties, especially the spatial diversity.
\end{enumerate}
To design a secure protocol based on physical-layer properties such as the spatial diversity, evaluating the mutual information between legitimate parties is crucial. It is important that sufficient and robust joint entropy exist. Due to the lack of space, we investigate an information theoretical evaluation of the empirical measures using:
\begin{enumerate}
    \item Mutual information estimator and
    \item Secret-key rate.
\end{enumerate}
We present a pre-processing cascade. However, we do not focus on key extraction protocols or aspects of quantization, information reconciliation or privacy amplification. For the interested reader we refer to~\cite{zenger_physical-layer_2017}.

\subsection{GNSS: A Global Network of 4 Independent Systems}
Important for a real-world system is global applicability, as well as trustworthiness. \gls{GNSS} are ideal candidates, because they are not governed by a single party, but multiple nations. Globally available systems are Beidou (China), Galileo (Europe), GLONASS (Russia), GPS (USA), with a total of 92 operational, multi-band capable satellites in \gls{MEO}. The combination of these systems leads to an average of 34 visible satellites anywhere on Earth~\cite{noauthor_navigation_nodate}. Beidou, QZSS (Japan) and IRNSS (India) provide additional satellites in geosynchronous orbits to further enhance the availability in specific regions~\cite{noauthor_list_nodate}. 

Polar regions suffer from less positioning accuracy, because satellites here only reach lower elevations~\cite{noauthor_challenges_nodate}. Urban and mountainous environments also limit the amount of available satellites, but the combination of multiple satellites ensures the availability of \gls{GNSS} signals practically anywhere outdoors. 

\subsection{Physical-Layer Properties}
\subsubsection{Spatial Channel Diversity and Randomness}
\begin{figure}[!t]
	\centering
	\includegraphics[width=\columnwidth]{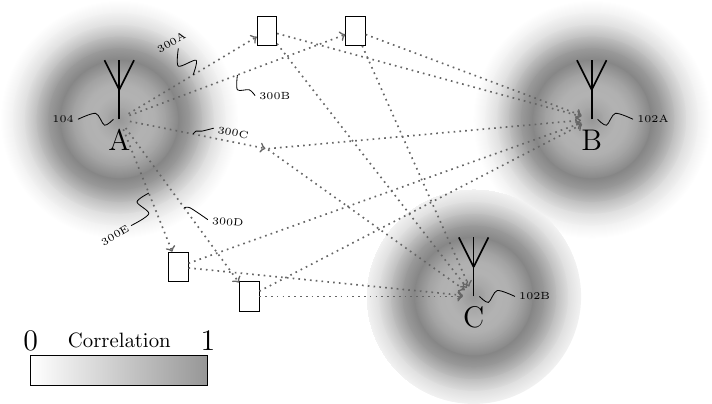}
	\caption{Major properties of the wireless channel: spatial diversity and randomness, which is a substitution of several radio waves with different propagation behavior. The sum results in a complex channel profile \cite{zenger_physical-layer_2017}.}
	\label{fig:channel_scd}
\end{figure}

Spatial channel diversity is crucial for assessing the security of key extraction schemes against adversaries. Besides two legitimate parties Alice $A$ and Bob $B$ in proximity, we add a third party Eve $E$ that is able to perform correlated channel observations. The strength of this correlation is a function of the environmental parameters, such as relative positioning between these three parties and the utilized wavelength for the radio communication. Generally, devices in close proximity sample highly correlated channel observations, whereas a decrease is visible with increasing distance, as shown in \Cref{fig:channel_scd}.

The wireless channel is a complex and dynamic environment due to many unpredictable multipath effects.  Each signal or multipath component can be represented by an amplitude and phase, typically denoted as a complex number with the sum resulting in a complex channel profile.

\subsubsection{Multipath Effects}
Besides the dominant line-of-sight component, the environmental interactions result in several non-line-of-sight components. These are known as multipath, assumed to be Rayleigh distributed, and can be summarized as follows \cite{zenger_physical-layer_2017}: shadowing (log-normal distributed) and absorption, reflections at large obstacles, refraction at medium crossing, scattering at small obstacles, and diffraction at edges.

Due to the high velocities of satellites in general, e.g.\ $\approx3.8$ km/s for \gls{GNSS} satellites in \gls{MEO}, the channel is time-varying due to the constantly changing relation of reflectors and scatterers to each other. This results in changing multipath components, hence in amplitude, frequency, and phase variations, where the latter are also related to the Doppler effect. 

\subsubsection{Ionospheric Effects} 
An important impact on radio signals in space communications is the ionosphere with large amounts of charged particles that are impacted by the sun's activity, the geomagnetic field, and the day/night cycle, with increased effects during the night and reduced ones during the day.  Electron density irregularities are caused by plasma instabilities and can lead to rapid small-scale changes, known as scintillations. Amplitude scintillations can result in a drop in signal level below a receiver's lock threshold, while phase scintillation can create cycle slips and additional noise that challenge a receiver's ability to hold lock on a signal \cite{sbas_ionospheric_working_group_effect_2010}. The former are dominant around the geomagnetic equator, the latter in polar regions. Most intense scintillations are found in the F-layer at a height of $150-500$ km \cite{orus_performance_2002}.

The speed of trans-ionospheric communications is reduced by the charged particles. Dual-frequency systems or correction data obtained from satellites or large-scale monitoring networks allow compensating this ionospheric delay. Even though large-scale ionospheric effects are correctable due to sophisticated empirical models, small-scale effects are unpredictable and result in global errors of $20-30\%$ \cite{orus_performance_2002}. 

\subsection{Channel Features and Pre-Processing} \label{preproc}
To extract joint entropy and create a universally applicable cryptography scheme, we use the channel properties of \gls{GNSS} links to our advantage. In essence, \gls{GNSS} receivers calculate their position by trilateration of the distances to at least four visible satellites, each of which broadcasts their exact position and time of transmission. The receiver calculates the distance to each satellite from the time delta. This calculated distance is called pseudorange, as it is affected by multiple error sources. These errors are the source of \gls{GNSS} positioning inaccuracies and include instrumental noise, clock errors, and effects that are exploited by us, i.e.\ ionospheric delay and multipath errors.

To enhance positioning accuracy, \gls{GNSS} satellites transmit at two frequencies simultaneously. Because the ionospheric delay and multipath effects are frequency dependent, dual-band \gls{GNSS} receivers are able to mitigate them through a linear combination of both signals~\cite{subirana_gnss_nodate}. We use the same principal to extract joint entropy based on small scale variations.

For each satellite, we combine the carrier phase measurements of both frequencies to calculate the delay. While carrier phase alone cannot be used for positioning applications due to the ambiguous number of full cycles during the signal propagation, they allow for a more accurate tracking of relative changes in pseudorange, as long as no cycle slip occurs.

Before the resulting time-series can be used for key establishment, further pre-processing steps are needed. Thus, we subtract a fitted polynomial from the time-series to remove predictable components, apply a Savitzky-Golay filter to remove instrumental noise and normalize the resulting data.

\subsection{Mutual Information}
The mutual information calculates the statistical dependency between two values $X$ (observation of Alice) and $Y$ (observation of Bob or Eve) with the formula shown in~(\ref{eq:mi}), where $p(\boldsymbol{\cdot})$ denotes the probability. 

\begin{equation}\label{eq:mi}
    I(X;Y)=\sum_{y \in Y}\sum_{x \in X}p(x,y) \cdot \log\left(\frac{p(x,y)}{p(x) \cdot p(y)}\right)
\end{equation}

We estimate the mutual information via a k-nearest neighbor principle developed by Kraskov et al. \cite{kraskov_estimating_2004}. The mutual information provides a metric for measuring the amount of information that we obtain from one variable when we observe the other in bit. It is upper bounded by the minimum of the entropy of the two variables. High mutual information between two vectors implies that a high amount of information is shared.
	
\subsection{Secret-Key Rate}
The secret-key rate $R_{sk}$ is defined as 

\begin{equation} \label{eq:secret_key_rate}
	R_{sk} := I(X;Y) - \text{min}[I(X;Z), I(Y;Z)]
\end{equation}

It represents the mutual information between two channel observations $X, Y$ while considering any information available to an eavesdropper with his observations $Z$, i.e.\ the mutual information $I(X;Z)$ and $I(Y;Z)$. A positive secret-key rate indicates that the observations of the legitimate parties share information that the attacker could not obtain \cite{zenger_physical-layer_2017}.

%% file: 3_results.tex
\section{Evaluation and Testbed}

We empirically evaluate the usability of satellite links for spatial channel diversity based PLS with a testbed built from off-the shelf GNSS receivers and use $R_{sk}$ as a main metric for our evaluation.

The central research question R0 we address is: how pronounced is the classic decorrelation behavior (over distance between $A$, $B$ and $A$, $E$), known from Jake's model \cite{jakes_microwave_1995, goldsmith_wireless_2005}, of satellite channels? We identified several 2\textsuperscript{nd}-order research questions which will be addressed and partially evaluated later:
\begin{itemize}
    \item[$\triangleright$] R1: What physical effects are essential for the randomness, i.e.\ ionosphere and multipath?
    \item[$\triangleright$] R2: Do the expected stronger ionospheric effects in north and south directions and the diurnal behavior of the ionosphere affect the observed entropy?
    \item[$\triangleright$] R3: What are useful criteria for choosing useful satellite positions (relative to $A$ and $B$)?
    \item[$\triangleright$] R4: Are there performance differences with respect to the GNSS constellation?
    \item[$\triangleright$] R5: Do GNSS provide a sufficiently high $R_{sk}$ for a key extraction anywhere on Earth?
\end{itemize}

The testbed consists of three ArduSimple simpleRTK2B v1 boards that are equipped with a ublox ZED-F9P \gls{GNSS} chip which is connected to a calibrated survey \gls{GNSS} multiband antenna. The signals of every visible satellite from any of the four GNSS and QZSS constellations are sampled at 20 Hz. IRNSS is not supported. Measurements are obtained as UBX data and stored for processing by a Raspberry Pi Zero. The testbed is battery-powered and can run for 32 hours.

\begin{figure}[!t]%
\centering
\subfloat[]{\label{image_roof}\includegraphics[width=(\columnwidth-0.05in)/2]{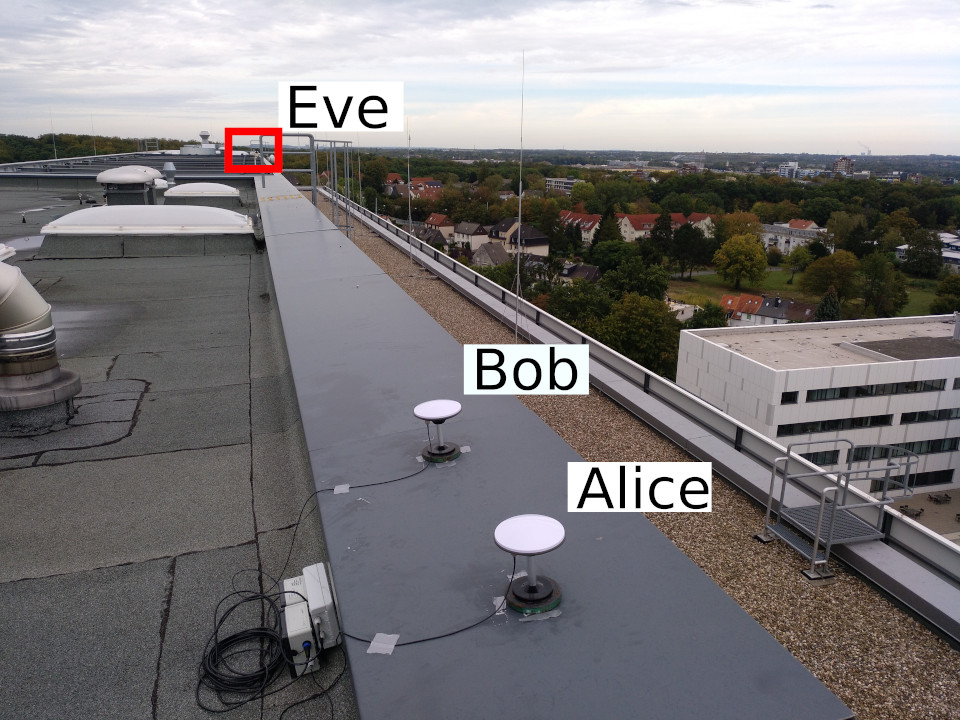}}\hfill
\subfloat[]{\label{image_courtyard}\includegraphics[width=(\columnwidth-0.05in)/2]{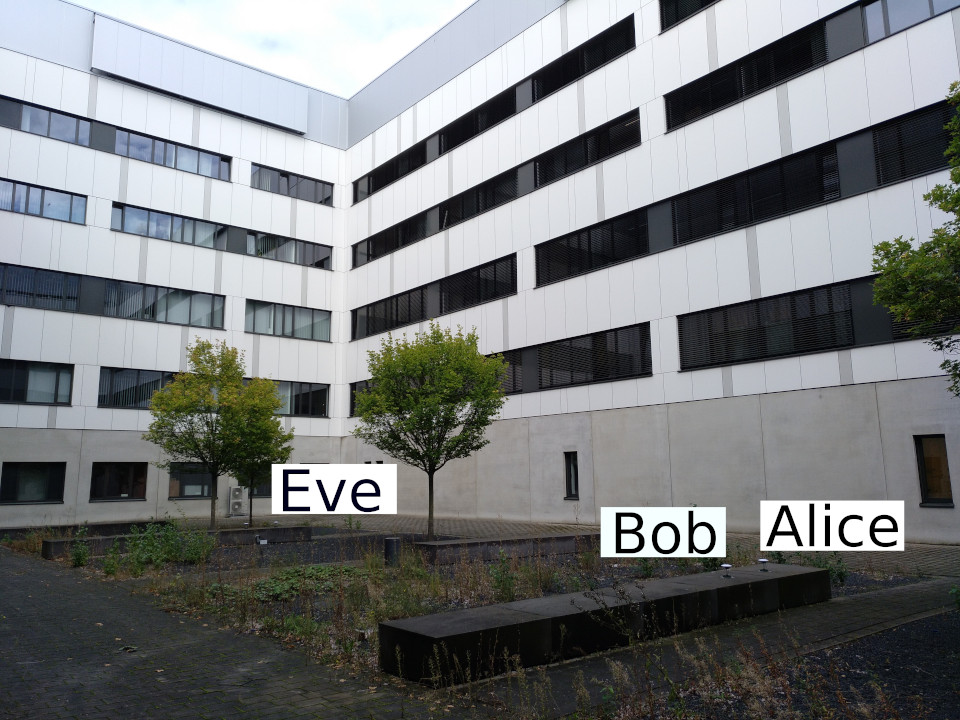}}%
\caption{Environments of the experiments. (a) Roof with reduced shadowing. (b) Courtyard with increased shadowing and reflections.}
\label{images_experiments}
\end{figure}

The experiments are conducted in two contrasting environments that are shown in \Cref{images_experiments}. Firstly, the top of a tall (12 floors) university building is chosen as an environment with reduced shadowing due to lacking obstructions in the surrounding area (cf. \Cref{image_roof}). $A$ and $B$ are positioned at a distance of 80 cm, whereas the attacker is placed at a distance of 75 m. Secondly, the inner courtyard of a university building is chosen as an environment with increased shadowing and reflections induced by the building, as well as scattering of nearby trees (cf. \Cref{image_courtyard}). $A$ and $B$ maintain a distance of 1 m, while the attacker is now placed 20 m distance due to spatial limitations.

We split the obtained measurements per satellite and 5 minute time frame. We omit all blocks that were not continually tracked by all receivers at two frequencies and without cycle slips. 
We choose the parameters for the pre-processing (cf.~\Cref{preproc}) as degree 5 for the polynomial and window size 81, order 1 for the Savitzky-Golay filter. $R_{sk}$ is estimated based on the resulting time-series of $6000$ samples.

\section{Evaluation Results}
\subsection{Representative Measurement of a Satellite}
\begin{figure}[!t]
	\centering
	\includegraphics[clip, width=\columnwidth]{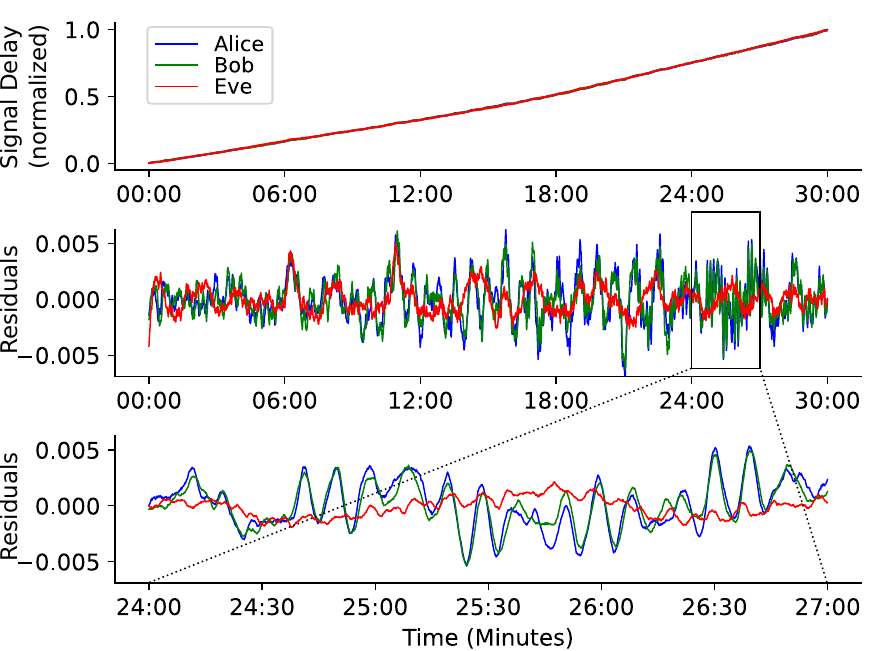}
	\caption{Measurements of $A$, $B$, and $E$ for satellite G27 in 30 min window, residuals ($R_{sk} = 0.93$) and closer look at a 3 minute window ($R_{sk} = 1.02$).}
	\label{repre_measurement}
\end{figure}

In order to evaluate R0, we expect to obtain highly correlated data between the legitimate parties, i.e.\ $A$ and $B$, whereas the eavesdropper $E$ cannot obtain equally correlated observations. \Cref{repre_measurement} shows a sample of the original measurements with a very close trend between all parties. The resulting residuals from the chosen polynomial fit show the differences between the observations. Focusing on a 3-minute window reveals highly correlated data between $A$ and $B$, with unrelated observations made by $E$. This illustrative example confirms our base hypothesis. There are however data blocks in our experiment that do not show the same decorrelation over distance and therefore do not confirm R0. 

\subsection{Analysis of all Satellites}
\begin{figure}[!t]%
    \centering
    \includegraphics[width=\columnwidth]{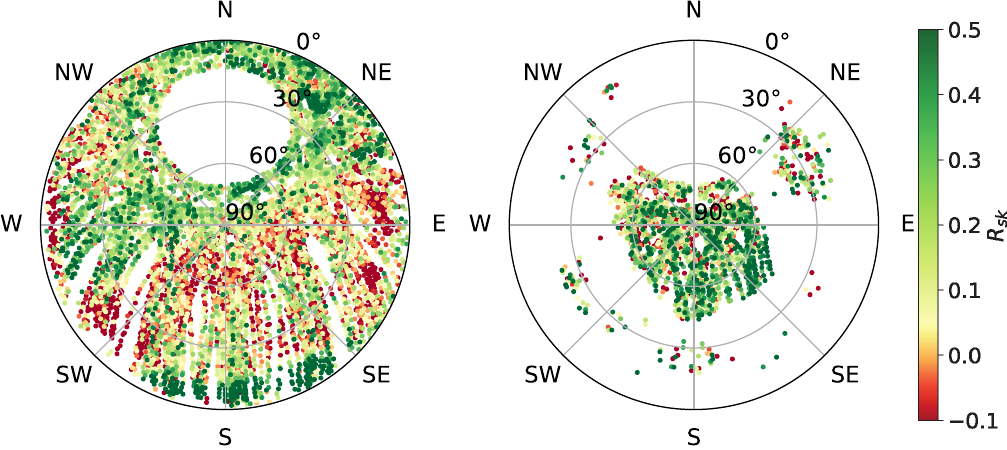}
    \caption{$R_{sk}$ for each satellite visible during the experiment on the roof (left) and courtyard (right).}
    \label{skyplots_rsk}
\end{figure}

For a better overview on the obtained data and the evaluation of the other hypotheses, we consider all captured satellites over the entire duration of the experiments. \Cref{skyplots_rsk} highlights the $R_{sk}$ for each 5 min block as a sky plot, which depicts satellite positions on the celestial sphere with elevation $\varphi$ from the horizon. The visible hole in coverage is related to the angle of the satellite orbits.

The left sky plot shows the results of the first experiment on the roof. Generally, we observe the highest $R_{sk}$ close to the horizon, with the largest values dominantly placed in the areas of NW--NE and SW--SE.  Lower and more negative values are present for higher elevations. Causes for the higher observed $R_{sk}$ in low elevations can be twofold. Firstly, in areas with few obstructions, multipath primarily affects signals received at low elevations, as these result in most interactions and reflections with the ground. Due to the longer signal path, the ionosphere is also interfering stronger with the signal. The areas of high $R_{sk}$ located in north and south can thus either be a result of the ionospheric delay, which would confirm R2, or a result of the geometry of the testing environment. We also considered a distribution of $R_{sk}$ over a period of 24 hours and did not find a clear effect of diurnal ionospheric changes on $R_{sk}$, not confirming this part of R2. 

The courtyard experiment, depicted in the right sky plot, logically does not show many data points in the lower elevations. Interestingly though, the $R_{sk}$ values at higher elevations are larger compared to the unobstructed experiment. In unobstructed scenarios, high elevations do not experience the aforementioned multipath effects at low elevations with Earth's surface, as we have a more dominant line-of-sight component. As a result, the courtyard provides sufficient multipath that even high elevations induce high $R_{sk}$.

Based on both observations, we can answer R1 and conclude that multipath is the largest contributor to high $R_{sk}$ in our experiment. Part of the lower ionospheric influence on $R_{sk}$ is likely due to the relatively small distance between the parties in our experiment and the resulting nearly identical \gls{IPP}. We expect increasing attacker distances to produce larger $R_{sk}$ even for signals with low multipath influence.

\begin{figure}[!t]%
    \centering
    \includegraphics[width=\columnwidth]{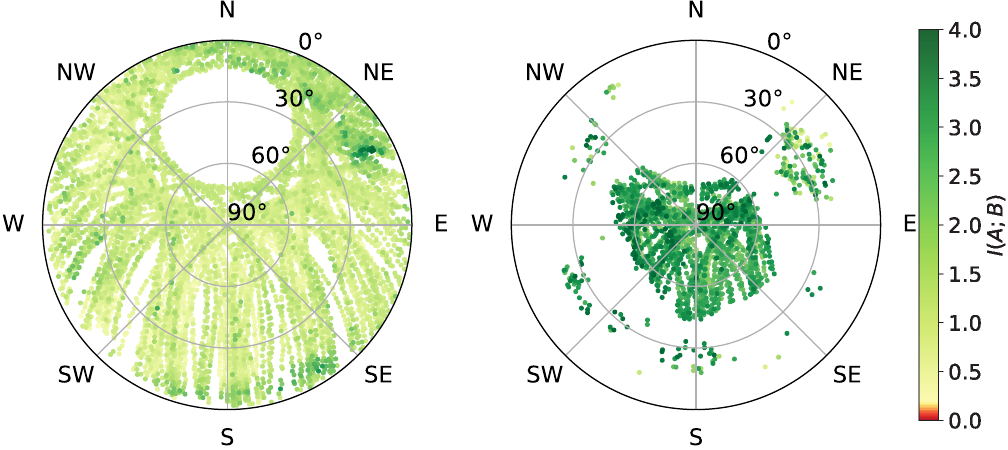}
    \caption{$I(A;B)$ for each satellite visible during the experiment on the roof (left) and courtyard (right).}
    \label{skyplots_miab}
\end{figure}

Another comparison of the available entropy in both environments is shown in \Cref{skyplots_miab}, which depicts the increased mutual information between $A$ and $B$ in the courtyard experiment, where the received signals are richer in information, typically above $2.5$ bit. In contrast, the roof experiments has an average maximum of around $1.0-1.5$ bit. This confirms the expected strong influence of multipath on observed mutual information. Although the courtyard experiment sometimes shows negative $R_{sk}$, indicating that $E$ observed the same information as the legitimate parties and is thus able to derive the same secret, one has to keep in mind the short distance of 20 m in our setup. An attacker in another environment is unlikely to observe the same patterns.

\begin{figure}[!t]
	\centering
	\includegraphics[width=\columnwidth]{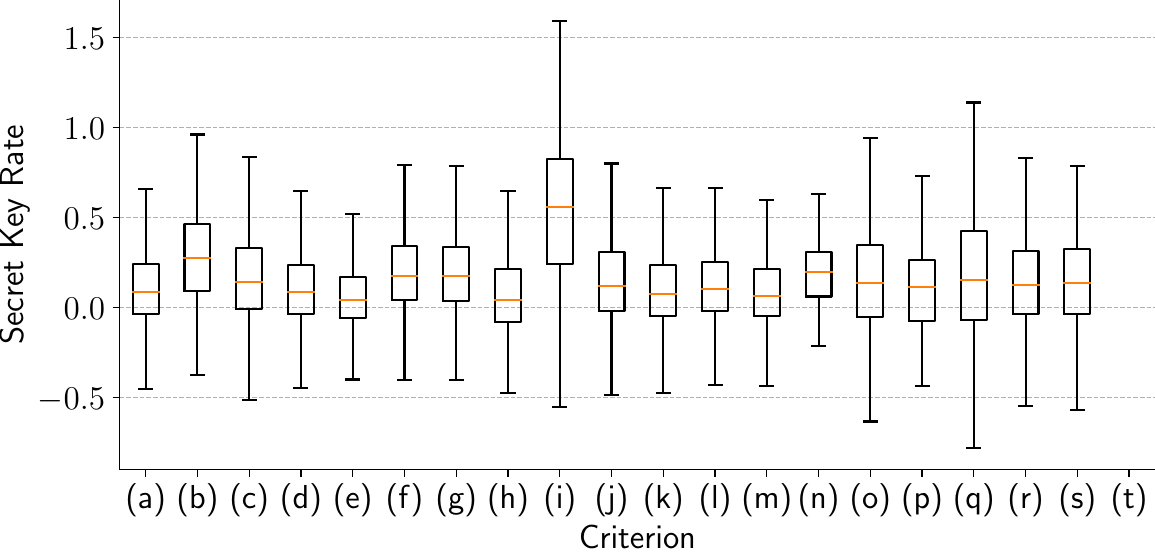}
	\caption{Distribution of $R_{sk}$ for various criteria.}
	\label{all_criteria}
\end{figure}

The observed negative $R_{sk}$ values show the importance of objective criteria to select usable satellites in real time, which is the focus of R3. We also consider possible differences in the usability of each GNSS constellation to evaluate R4. In order to partially answer these research questions, we present a quantitative analysis for both experiments in \Cref{tab:quant-overview}. For each criterion, we compare the average number of available satellites above different threshold $R_{sk}$ to the number of satellites with negative $R_{sk}$. The $R_{sk}$ distributions for these criteria are shown in \Cref{all_criteria}.

Notable criteria include (b) and (i), which lead to a particularly low chance of negative $R_{sk}$, but due to their strictness also limit the number of available satellites with positive $R_{sk}$ to less than 1 on average. We omit the satellite position analysis for the courtyard experiment, as available satellites are already situated in a very small window.

Filtering the data for separate constellations answers R4. All constellations seem equally likely to produce usable observations. QZSS satellites are rarely observed, so claims concerning those would need more data points.

Lastly, we see in \Cref{tab:quant-overview_min_availability} that almost for the entire time of both experiments, except only two 5 minute blocks, we make observations that result in a positive $R_{sk}$. As a result, R5 can be confirmed at least for our two experiments.

\begin{table}[!t]
    \centering
    \caption{Average number of satellites within criteria above a given threshold $R_{sk}$ compared to negative $R_{sk}$ over a period of 32h.}
    \label{tab:quant-overview}
    \begin{tabular}{cccccccc}\toprule
    & & Criterion / $R_{sk}$                     & $> 0.4$ & $> 0.2$ & $> 0$ & $\leq 0$ \\\midrule
    \parbox[t]{2mm}{\multirow{14}{*}{\rotatebox[origin=c]{90}{Roof}}}
    & (a) & All Data                               &  3.3    &  8.4    & 18.4  &  8.7 \\\cmidrule(l{.5em}){2-7}
    & (b) & $\varphi \leq 2\degree$                &  0.3    &  0.6    &  0.8  &  0.1 \\
    & (c) & $3\degree \leq \varphi \leq 10\degree$ &  0.9    &  1.9    &  3.4  &  1.3 \\
    & (d) & $10 < \varphi \leq 45\degree$          &  1.7    &  4.5    &  10   &  4.7 \\
    & (e) & $\varphi > 45\degree$                  &  0.4    &  1.5    &  4.2  &  2.6 \\\cmidrule(l{.5em}){2-7}
    & (f) & NW--NE                                 &  1.3    &  3.2    &  5.8  &  1.2 \\
    & (g) & NW--NE, $\varphi \leq 10\degree$       &  0.6    &  1.4    &  2.6  &  0.6 \\
    & (h) & SW--SE                                 &  0.7    &  1.3    &  3    &  2.1 \\
    & (i) & SW--SE, $\varphi \leq 10\degree$       &  0.4    &  0.5    &  0.6  &  0.1 \\\cmidrule(l{.5em}){2-7}
    & (j) & Beidou                                 &  0.4    &  1      &  1.8  &  0.7 \\
    & (k) & Galileo                                &  1.2    &  2.6    &  5.8  &  3   \\
    & (l) & GLONASS                                &  0.9    &  2.7    &  5.6  &  2.3 \\
    & (m) & GPS                                    &  0.7    &  2      &  4.9  &  2.6 \\
    & (n) & QZSS                                   &  0      &  0.1    &  0.2  &  0   \\\midrule
    \parbox[t]{2mm}{\multirow{6}{*}{\rotatebox[origin=c]{90}{Courtyard}}}
    & (o) & All Data                               &  1.3    &  2.5    &  4.3  &  1.9 \\\cmidrule(l{.5em}){2-7}
    & (p) & Beidou                                 &  0.1    &  0.2    &  0.3  &  0.2 \\
    & (q) & Galileo                                &  0.6    &  1      &  1.4  &  0.7 \\
    & (r) & GLONASS                                &  0.3    &  0.7    &  1.3  &  0.5 \\
    & (s) & GPS                                    &  0.3    &  0.7    &  1.3  &  0.5 \\
    & (t) & QZSS                                   &  0      &  0      &  0    &  0   \\\bottomrule
    \end{tabular}
\end{table}

\begin{table}[!t]
    \centering
    \caption{Availability of at least 1 satellite with $R_{sk}$ above a given threshold over a period of 32h for both experiments.}
    \label{tab:quant-overview_min_availability}
    \begin{tabular}{ccccccc}\toprule
    Criterion / $R_{sk}$                     & $> 0.4$ & $> 0.2$ & $> 0$ & $\leq 0$ & $\leq -0.2$ \\\midrule
    Secure bits / second                     & $> 8$   & $4-8$   & $0-4$ & 0        &  0          \\\midrule
    \multirow{2}{*}{Roof}                    & 30 h    & 31.9 h  & 31.9 h& 31.9 h   & 23.7 h      \\
                                             & 93.8\%  & 99.7\%  & 99.7\%& 99.7\%   & 74.0\%      \\
    \multirow{2}{*}{Courtyard}               & 23.6 h  & 30 h    & 31.9 h& 28 h     & 16.4 h      \\
                                             & 73.7\%  & 93.8\%  & 99.7\%& 87.5\%   & 51.3\%      \\\bottomrule
    \end{tabular}
\end{table}

%% file: 4_discussion_future_work.tex
\section{Discussion and Future Work}

We have shown the feasibility of GNSS signals as basis for PLS. More experiments will expand our results to different environments. Polar regions demand further focus, as they often suffer from lower GNSS performance due to satellites only reaching lower elevations~\cite{noauthor_challenges_nodate}. As low elevation satellites have been shown to be more reliable sources of entropy, this may not at all present as a disadvantage for our proposal.

We conclude from our observations that the potential of multipath effects to produce a high secret-key rate is larger than that of the ionosphere. The ionospheric impact is likely to increase with larger distances, so an empirical evaluation of such scenarios is useful. A frequency analysis of data in different scenarios will give further insights on the characteristic of effects related to multipath and ionosphere.

Future analyses will also pinpoint the relative magnitude of mutual information available to the legitimate and eavesdropping parties to derive ``secure perimeters'' and solidify criteria legitimate parties can use to determine satellites providing positive secret-key rates.

The pre-processing of the data can be further optimized to the next steps of the key-generation process, i.e.\ quantization, information reconciliation, and privacy amplification. This is also an indicator for the usability for different kinds of applications, due to the varying time-slots available for obtaining signals. As we obtained already promising results for 5 minute blocks of data, further analyses for smaller time-frames can be conducted. 

Lastly, the availability of global modeling and prediction of the ionosphere has to be considered as a security critical side-channel. However, as these maps are typically available at a temporal resolution of 15 minutes and report discrete values in a grid of several degrees in latitude and longitude, they have a negligible advantage compared to the conducted high-rate measurements.

\section{Conclusion}
In this paper, we introduce the idea and evaluation of physical-layer security using satellite-to-Earth communication. By extending a known \gls{PLS} approach from the terrestrial context to \gls{GNSS} signals, we improved its security and applicability. The approach might have an impact to future privacy-preserving proximity verification protocols.

We have implemented a functional prototype using three off-the-shelf hardware testbeds simulating the parties Alice, Bob, and Eve. The general feasibility based on two experiments over 32 hours is evaluated, with signals obtained from 81 and 77 unique satellites, respectively. Our obtained datasets are available for further analyses\footnote{DOI: 10.5281/zenodo.4084857}. 

We identified the phase difference of dual-frequency \gls{GNSS} satellites as a suitable parameter. Then we analyzed the mutual information between the channel observations of Alice and Bob, as well as the secret-key rate by using the observations of Eve. Our preliminary results demonstrated both positive and negative rates and motivated us to dive deeper into understanding the underlying reasons. By splitting the data into different groups, i.e., elevation, geographical quadrant, and \gls{GNSS} system, we were able to create reasoned origin hypotheses.

\section*{Acknowledgements}
This work was funded in part by the German Federal Ministry of Education and Research (BMBF) (Grant KIS6KIITS10 mINDFUL) and by the German Research Foundation~(DFG) within the framework of the Excellence Strategy of the Federal Government and the States - EXC2092 CASA - 390781972.